# Determination of the Nature of the Tetragonal to Orthorhombic Phase Transition in SrFe$_2$As$_2$ by Measurement of the Local Order Parameter

J.C. Loudon[1]*, C.J. Bowell[1], J. Gillett[2], S.E. Sebastian[2], P.A. Midgley[1]

[1]*Department of Materials Science and Metallurgy, University of Cambridge, Pembroke Street, Cambridge, CB2 3QZ, United Kingdom.*

[2]*Department of Physics, University of Cambridge, Cavendish Laboratory, JJ Thomson Avenue, Cambridge CB3 0HE, United Kingdom.*

SrFe$_2$As$_2$ is the end-member for a series of iron-pnictide superconductors and has a tetragonal-to-orthorhombic phase transition near 200 K. Previous macroscopic measurements to determine the nature of the transition gave seemingly inconsistent results so we use electron microscopy to monitor the local order parameter showing that the transformation is first order and that the orthorhombic phase grows as needle domains. This suggests the transition occurs via the passage of transformation dislocations, explaining the apparent inconsistencies. This mechanism may be common to similar transitions.

PACS numbers: 64.70.K-, 74.70.-b

SrFe$_2$As$_2$ has been the subject of recent investigation as it is the parent compound for a class of the newly discovered iron-pnictide superconductors. SrFe$_2$As$_2$ itself does not superconduct at ambient pressure but much attention has been paid to the tetragonal-to-orthorhombic phase transition it exhibits near 200 K as it is associated with the formation of a static spin-density wave [1]. When SrFe$_2$As$_2$ is doped, e.g. with potassium, the ground state becomes superconducting rather than a spin-density wave [1]. The connection between these ground states is regarded as an important clue to the superconducting mechanism [2].

Whenever a phase transition is described, the question almost always arises as to whether it is first order in nature, involving the coexistence of two phases; or second order, where one phase changes uniformly and continuously into another [3]. This distinction appears clear-cut but there is frequently disagreement over how transitions should be classified. Two examples are the paramagnetic-to-ferromagnetic transition in colossal magnetoresistive manganites [4] and even the cubic-to-tetragonal in SrTiO$_3$, often cited as a prototypical second order phase transition, is difficult to classify [5]. Similar disagreements surround the structural phase transition in SrFe$_2$As$_2$.

It seems natural to use a microscopic technique to determine the order of any phase transition as imaging two coexisting phases would show at once that it is first order and yet this is not normally done: they are usually classified on the basis of bulk measurements such as heat capacity, x-ray or neutron diffraction and magnetisation. We believe this is part of the problem in determining the order of a phase transition and so introduce a technique for measuring the local order parameter.

A second difficulty is that it is possible to identify a first order transition but not to prove that a transition is second order: only that within the resolution of an experiment, the order parameter is not seen to change abruptly leaving the possibility that a more sensitive experiment will detect a jump invisible to previous measurements. These difficulties can be circumvented if the mechanism by which the transition takes place is described: here this is the process by which the atoms rearrange themselves. The order of the transition is then known automatically. From the microscopic measurements we make here, it is possible to do this.

Figure 1 shows the crystal structure of SrFe$_2$As$_2$ determined by Tegel *et al.* using x-ray diffraction [2]. The high temperature tetragonal phase has space group *I*4/*mmm* with lattice parameters $a_T = b_T = 3.92$ Å, $c_T = 12.36$ Å and the low temperature orthorhombic phase has space group *Fmmm* with $a_O = 5.58$ Å, $b_O = 5.52$ Å, $c_O = 12.30$ Å at 90 K. The distortion by which one structure is turned into another is a pure shear when viewed down the **c** axis: the right angle between **a**$_T$ and **b**$_T$ is reduced to 89.3° at 90 K in the orthorhombic structure. We define the order parameter, $Q$, for the phase transition as the 'orthorhombicity', $Q = \dfrac{a_o - b_o}{a_o}$.

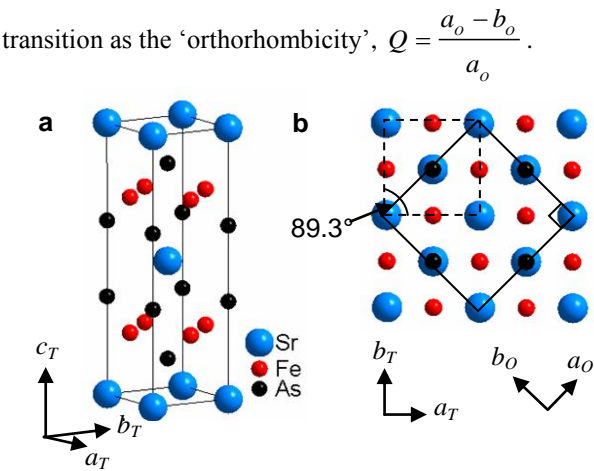

**Figure 1** (a) Tetragonal and (b) Orthorhombic SrFe$_2$As$_2$ viewed down [001] showing the relationship between the tetragonal (dashed line) and orthorhombic unit cells (solid lines). (Data from ref. 2.)

This change in space-group can be first or second order. Using x-ray diffraction of polycrystalline samples, Tegel *et al.* [2] found the transition was second order but Jesche *et al.* [6] found it was first order and both gave transition temperatures of 205 K. Yan *et al.* [7] made similar measurements on single crystals and one might think that a sharper transition would be obtained. Instead, they found phase coexistence in the temperature range 160–198 K and describe it as having a 'more complex nature'. By measuring the local order parameter we show that the phase transition is first order and classify the atomic rearrangement mechanism as martensitic enabling an explanation of these apparent discrepancies.

We used single crystals of $SrFe_2As_2$ from the batch used in ref. 8 grown by the Sn flux method. Electron microprobe analysis put an upper limit on tin inclusion of 0.3% Sn:Sr. The crystal structure is layered and samples were prepared for electron microscopy by repeated cleaving by placing the sample between two sheets of sticky-tape and pulling them apart. This yielded samples with large (~10 μm) electron transparent regions near their edges which were placed in clam-shell copper grids. The sample from which the large angle convergent beam diffraction patterns were taken was ion thinned after cleaving giving a very large (~30 μm) thin area. The ion thinning was performed using a Gatan Precision Ion Polishing System operated at 3 kV for half an hour until a small hole appeared in the sample. The results reported here could be observed in samples made with and without ion-thinning.

Images were taken using imaging and photographic plates and TV rate cameras using Philips CM30 and CM300 transmission electron microscopes. The samples were cooled using a Gatan liquid-nitrogen-cooled specimen stage and the temperatures quoted are measured by a thermocouple at the end of the holder. Previous experiments on phase transitions using the same holder indicated that the temperature is within 10 K of the true specimen temperature (see ref. 9). In the videos, the time at which the temperature changed by 1 K was noted and the temperatures quoted in still frames are extrapolated from this.

When the orthorhombic structure forms, there are four distinct orientations which were equivalent in the tetragonal phase so a twinned structure occurs with two twin variants having $(110)_O$ twin boundaries and two having $(\bar{1}10)_O$ boundaries. Figure 2(a) shows a schematic of the tetragonal unit cell and (b) shows two twin variants of the orthorhombic cell. The expected electron diffraction pattern is shown below each. Twinning causes a splitting of the Bragg reflections and the splitting angle, $\theta$, is used to measure the order parameter in the x-ray diffraction measurements we refer to and the microscopy techniques we use as $Q = \theta/2$ for small $\theta$.

Figure 2(c)-(f) show bright field images taken as a video on warming ([Supp. Info. 1](#)) from a region containing two orthorhombic twin variants. The specimen was oriented so that one set of twins was diffracting strongly and appears dark. It can be seen that one set of twins withdraws and there are needle tips at the end of the twins. It is tempting to view the twins as belonging to the orthorhombic phase and the phase left behind as tetragonal but this may not be the case. It is possible that the stresses the material experiences during warming cause the twins to withdraw leaving behind an untwinned orthorhombic region and the phase transition takes place

at a higher temperature. The fact that the withdrawal of twins in this region took place at 147 K, considerably below the transition temperature of 198 K shown by the sudden downturn in the plot of magnetisation versus temperature (see [Supp. Info. 2](#)), seems to add weight to this argument. In other regions of sample, twins withdrew at temperatures ranging from 115 to 182 K.

An interesting feature can be seen in Figure 2: even after the twins have withdrawn, evidence that they are still there can be seen by the jump in the position of the bend contours in (e), indicated by arrows. This effect can no longer be seen in (f). We can suggest what is happening based on the measurement of the local order parameter described later. The withdrawal of twins is in fact the phase transition. First one set of twins withdraws leaving behind the tetragonal phase coexisting with the opposing set of orthorhombic twins. About 2 K higher, the other set of orthorhombic domains withdraws. The suppressed transition temperature is due to the martensitic phase transition, described later.

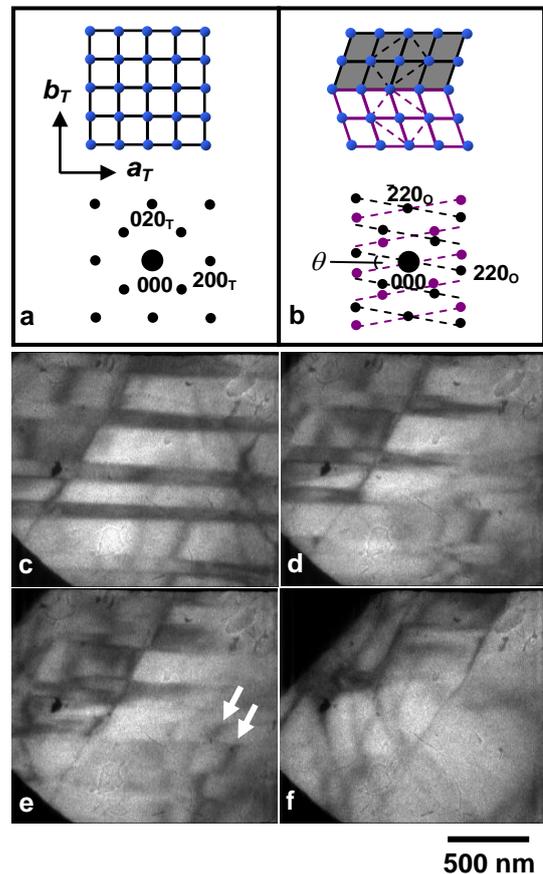

**Figure 2.** (a) Schematic of the tetragonal unit cell and its electron diffraction pattern. Note that reflections of type $h + k + l =$ odd are forbidden by the space group. (b) Schematic showing two twin variants of the orthorhombic cell (dashed lines) and its diffraction pattern. (c)-(f) Stills from a video of bright field images taken on warming (see text). (c) 146.1 K, 0 s, (d) 146.7 K, 17.0 s (e) 147.0 K, 24.4 s, (f) 148.5 K, 67.8 s.

Since it is not clear that one can infer the growth of a new phase from the growth of twins, we measured the spatially resolved order parameter using large angle convergent beam electron diffraction (LACBED) which combines real and reciprocal space information. Midgley *et al.* used this method to measure the local orthorhombicity of twins in $YBa_2Cu_3O_{7-\delta}$ [10]. The electron beam is converged so every part of the sample receives electrons from a different angle (see Figure 3(a)). Where the electron wave impinges at the Bragg angle for a particular set of planes, a dark contour is seen. The orientation of the atomic planes associated with each twin type is slightly different and the position of the dark contour jumps on moving from one twin to another. The separation of these contours is proportional to the order parameter as explained in Supp. Info. 3(A).

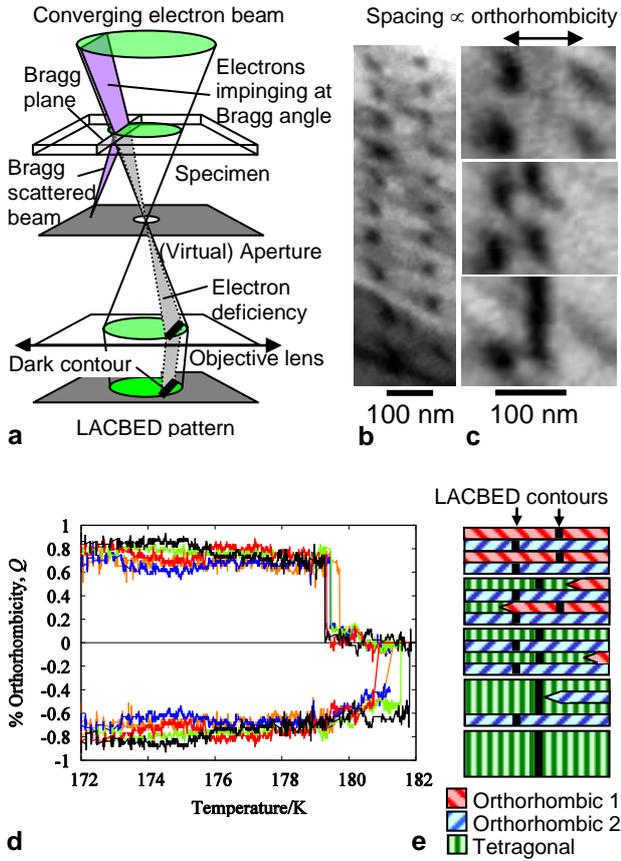

**Figure 3.** (a) Schematic explaining LACBED (see text). (b) LACBED pattern taken at 172 K, prior to the transition. (c) Enlarged stills from a video showing LACBED patterns taken on warming through the transition. The order parameter is proportional to the spacing between the contours. (d) The orthorhombicity, Q, versus temperature extracted from the video. Each colour represents a different pair of twins. (e) Schematic illustrating the phase transition and the positions of the LACBED contours.

LACBED patterns were recorded as a video on warming in a region containing twenty twins (Figure 3(b), for video see Supp. Info. 4) and enlargements showing the stages of the transition are shown in Figure 3(c). Two series of jumps in the position of the LACBED contours near the transition were observed. In the first, the dark contours on the right jump one at a time to the left whilst the ones on the left remain unchanged. Each jump is faster than the time between video frames (1/25 s). The distance between the two sets of contours after the jump is halved showing that one set of twins has transformed to the tetragonal phase but the opposing twins remain orthorhombic. About 2 K higher, the second set of dark contours on the left jump to the right whereas those on the right remain unchanged. The LACBED contour is now unbroken showing that the region of interest is all tetragonal.

Figure 3(d) was constructed by measuring the separation of elements of the broken LACBED contour between neighbouring twins as explained in Supp. Info. 3(B) and (C). Positive values of the order parameter refer to one set of twins and negative values to the other.

The LACBED video demonstrates that the transition is first order with one orthorhombic twin type becoming tetragonal one twin at a time, followed by the other. Unless each twin instantly becomes tetragonal throughout its entire length, each orthorhombic twin must withdraw along its length, leaving behind the tetragonal phase. This is what Figure 2 showed, enabling the description of the transition shown in Figure 3(e). All the twins are drawn withdrawing in the same direction as appears to be the case in Figure 2 but it is possible that some twins withdraw in the opposite direction. LACBED videos taken on cooling (Supp. Info. 5) show that the tetragonal-to-orthorhombic transition is the same process in reverse. It should also be mentioned that if the temperature is held constant, the transition stops midway through and remains stable. We spent some time searching for the 'tweed' microstructure, often regarded as a precursor to a structural phase transition [11], but this was not observed.

The phase transition takes place via the withdrawal of orthorhombic twins along their length on warming leaving behind the tetragonal phase and the reverse on cooling and we now discuss the interface between the two phases. Figure 2(c)-(f) showed that the orthorhombic twins ended in needle tips as they withdrew. This is reminiscent of the 'needle twins' (also called 'lenticular twins') observed in ceramics and metals where one twin variant ends and the other begins. In materials where the atomic structure of the needle tip has been observed, it is found that the boundaries are a series of atomic steps and that at each step there is necessarily a twinning dislocation, shown schematically in Figure 4(a). Twinning dislocations are edge dislocations with the twin boundary being the slip plane and the Burgers vector, **B**, in the slip plane and pointing in the direction of the needle tip with a magnitude given approximately by $|\mathbf{B}| = \sqrt{2}Qa_o$, a small percentage of the lattice vector. The subject is reviewed in ref. 12 and this structure has been observed directly by high resolution transmission electron microscopy in ceramics as diverse as $YBa_2Cu_3O_{7-\delta}$ [13] and $PbTiO_3$ [14] and used to model needle tips in simple metals since the 1950s [15].

Twins grow or shrink via the movement of twinning dislocations and we suggest that as the tetragonal-to-orthorhombic transition closely resembles the growth of twins, the process should have a similar explanation. Figure 4(b) shows the proposed structure of a needle domain of the orthorhombic phase penetrating the tetragonal phase. At each step in the needle domain, there is a *transformation dislocation* and when these move along their glide planes, one phase is transformed into another. Transformation dislocations are almost identical to twinning dislocations except that the magnitude of their Burgers vector is halved. This type of transition, occurring through the generation and movement of transformation dislocations, is termed a martensitic phase transition [11]. An athermal martensitc transition is one where the extent of the transformation depends on temperature not on time and from the observation that the phase transition can be stopped midway through by holding the temperature constant, it appears that this is what we have here. We have shown that the needle domains advance or retreat in a series of jerks which we interpret as the dislocations jumping from one pinning site to the next. A sufficient degree of undercooling is needed to provide the driving force to cause the pinning to be overcome which is why the partially transformed state remains if the temperature is held constant.

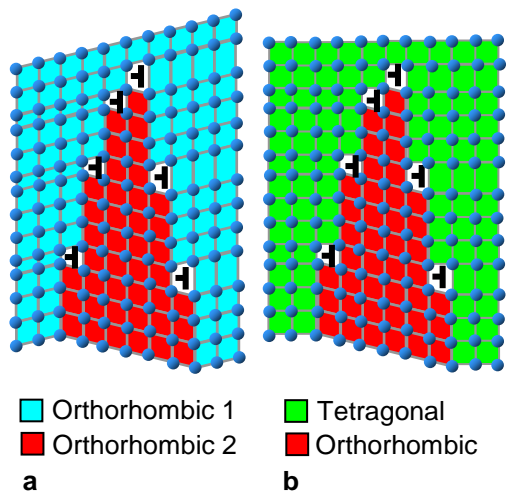

**Figure 4** (a) The staircase structure of a needle tip where one twin type (orthorhombic 1) meets another (orthorhombic 2). (b) The proposed coexistence of the tetragonal and orthorhombic phases (see text).

Transformation dislocations are most likely to be generated at defective regions of the sample or at the surfaces. To form a needle domain, it is necessary to generate many dislocations, one on each adjacent atomic plane. This seems unlikely to occur by chance and Cottrell and Bilby have described a mechanism similar to the Frank-Read source of dislocations whereby the movement of one dislocation can generate further dislocations on neighbouring planes [15]. A repetition of this process produces the needle shaped tip and once dislocation multiplication ends, a parallel-sided domain with a needle tip is formed. This model is not entirely satisfactory as it requires a fortuitous set of circumstances to be satisfied at once, described in ref. 15. It is also unclear why regularly spaced orthorhombic domains of one twin variant are generated first on cooling and initially coexist with the tetragonal phase and the other twin variant occurs at a lower temperature. It appears further research is required.

We can now explain why x-ray diffraction measurements show sharp transitions in polycrystals but diffuse transitions with phase coexistence over almost 40 K are observed in single crystals. It is simply that on cooling, transformation dislocations can form at grain boundaries in the polycrystal but these nucleation sites are unavailable in the single crystal so the new phase forms at a much smaller number of defects and on the sample surfaces. It requires greater undercooling to drive the new phase the greater distances from the nucleation sites into the bulk and so the partially transformed state remains over a wider range of temperatures. From the observation that on warming, the transition appears to be the reverse of cooling, it seems the transformation dislocations follow the same paths they did on cooling but in reverse. As they move, they encounter the same pinning sites as on cooling and the same temperatures are required to supply the energy to overcome these obstacles leading to a similar range of phase coexistence temperatures on warming as on cooling. It should be noted that the coexistence region may be much larger than 40 K as one neutron diffraction study found that the orthorhombic phase could still be detected as high as 450 K, some 250 K above the transition temperature [16].

Usually diffuse transitions are taken as a sign of impurities in the sample but here it is the purer sample that shows the more diffuse transition. The spread of transition temperatures was not due to Sn incorporation as the images of the growth of the new phase clearly showed that the domains grew along their length and not as an array of patches with different transition temperatures. The diffuse transition is a result of the martensitic nature of this first order phase transition which occurs despite the space group symmetry of the two phases allowing a second order transition. We suggest that the transformation mechanism described here may be common to many other similar phase transitions.

This work was funded by the Royal Society and the EPSRC.

*Corresponding author. Email: j.c.loudon@gmail.com